
%
%
\catcode`\@=11 
\newcount\yearltd\yearltd=\year\advance\yearltd by -1900
%

\def\draftmode{\message{ DRAFTMODE }\def\draftdate{{\rm preliminary draft:
\number\month/\number\day/\number\yearltd\ \ \hourmin}}%
\headline={\hfil\draftdate}\writelabels\baselineskip=20pt plus 2pt minus 2pt
 {\count255=\time\divide\count255 by 60 \xdef\hourmin{\number\count255}
  \multiply\count255 by-60\advance\count255 by\time
  \xdef\hourmin{\hourmin:\ifnum\count255<10 0\fi\the\count255}}}
\def\nolabels{\def\wrlabeL##1{}\def\eqlabeL##1{}\def\reflabeL##1{}}
\def\writelabels{\def\wrlabeL##1{\leavevmode\vadjust{\rlap{\smash%
{\line{{\escapechar=` \hfill\rlap{\sevenrm\hskip.03in\string##1}}}}}}}%
\def\eqlabeL##1{{\escapechar-1\rlap{\sevenrm\hskip.05in\string##1}}}%
\def\reflabeL##1{\noexpand\llap{\noexpand\sevenrm\string\string\string##1}}}
\nolabels
%
\global\newcount\secno \global\secno=0
\global\newcount\meqno \global\meqno=1
\def\newsec#1{\global\advance\secno by1\message{(\the\secno. #1)}
\global\subsecno=0\eqnres@t\noindent{\bf\the\secno ~#1}
\writetoca{{\secsym} {#1}}\par\nobreak\medskip\nobreak}
\def\eqnres@t{\xdef\secsym{\the\secno.}\global\meqno=1\bigbreak\bigskip}
\def\sequentialequations{\def\eqnres@t{\bigbreak}}\xdef\secsym{}
\global\newcount\subsecno \global\subsecno=0
\def\subsec#1{\global\advance\subsecno by1\message{(\secsym\the\subsecno. #1)}
\ifnum\lastpenalty>9000\else\bigbreak\fi
\noindent{\it\secsym\the\subsecno ~#1}\writetoca{\string\quad
{\secsym\the\subsecno.} {#1}}\par\nobreak\medskip\nobreak}
\def\appendix#1{\global\meqno=1\global\subsecno=0\xdef\secsym{\hbox{#1}}
\bigbreak\bigskip\noindent{\bf #1}
\writetoca{{#1}}\par\nobreak\smallskip\nobreak}
%
%
\def\eqnn#1{\xdef #1{(\secsym\the\meqno)}\writedef{#1\leftbracket#1}%
\global\advance\meqno by1\wrlabeL#1}
\def\eqna#1{\xdef #1##1{\hbox{$(\secsym\the\meqno##1)$}}
\writedef{#1\numbersign1\leftbracket#1{\numbersign1}}%
\global\advance\meqno by1\wrlabeL{#1$\{\}$}}
\def\eqn#1#2{\xdef #1{(\secsym\the\meqno)}\writedef{#1\leftbracket#1}%
\global\advance\meqno by1$$#2\eqno#1\eqlabeL#1$$}
%
%
\global\newcount\refno \global\refno=1
\newwrite\rfile
\def\ref{$^{\the\refno}$\nref}
\def\nref#1{\xdef#1{\the\refno.}\writedef{#1\leftbracket#1}%
\ifnum\refno=1\immediate\openout\rfile=refs.tmp\fi
\global\advance\refno by1\chardef\wfile=\rfile\immediate
\write\rfile{\noexpand\item{#1\ }\reflabeL{#1\hskip.31in}\pctsign}\findarg}
\def\findarg#1#{\begingroup\obeylines\newlinechar=`\^^M\pass@rg}
{\obeylines\gdef\pass@rg#1{\writ@line\relax #1^^M\hbox{}^^M}%
\gdef\writ@line#1^^M{\expandafter\toks0\expandafter{\striprel@x #1}%
\edef\next{\the\toks0}\ifx\next\em@rk\let\next=\endgroup\else\ifx\next\empty%
\else\immediate\write\wfile{\the\toks0}\fi\let\next=\writ@line\fi\next\relax}}
\def\striprel@x#1{} \def\em@rk{\hbox{}}
\def\lref{\begingroup\obeylines\lr@f}
\def\lr@f#1#2{\gdef#1{\ref#1{#2}}\endgroup\unskip}

\def\addref#1{\immediate\write\rfile{\noexpand\item{}#1}} 
\def\startrefs#1{\immediate\openout\rfile=refs.tmp\refno=#1}
\def\xref{\expandafter\xr@f}\def\xr@f#1.{#1}
\def\cite{\expandafter\cxr@f}\def\cxr@f#1.{$^{#1}$}
\def\xcite{\expandafter\xcxr@f}\def\xcxr@f#1.{{#1}}
\def\refs#1{\count255=1$^{\r@fs #1{\hbox{}}}$}
\def\r@fs#1{\ifx\und@fined#1\message{reflabel \string#1 is undefined.}%
\nref#1{need to supply reference \string#1.}\fi%
\vphantom{\hphantom{#1}}\edef\next{#1}\ifx\next\em@rk\def\next{}%
\else\ifx\next#1\ifodd\count255\relax\xref#1\count255=0\fi%
\else#1\count255=1\fi\let\next=\r@fs\fi\next}
\newwrite\lfile
{\escapechar-1\xdef\pctsign{\string\%}\xdef\leftbracket{\string\{}
\xdef\rightbracket{\string\}}\xdef\numbersign{\string\#}}

\def\writestop{\def\writestoppt{\immediate\write\lfile{\string\pageno%
\the\pageno\string\startrefs\leftbracket\the\refno\rightbracket%
\string\def\string\secsym\leftbracket\secsym\rightbracket%
\string\secno\the\secno\string\meqno\the\meqno}\immediate\closeout\lfile}}
\def\writestoppt{}\def\writedef#1{}
\def\seclab#1{\xdef #1{\the\secno}\writedef{#1\leftbracket#1}\wrlabeL{#1=#1}}
\def\subseclab#1{\xdef #1{\secsym\the\subsecno}%
\writedef{#1\leftbracket#1}\wrlabeL{#1=#1}}
\newwrite\tfile \def\writetoca#1{}
\def\leaderfill{\leaders\hbox to 1em{\hss.\hss}\hfill}
\def\writetoc{\immediate\openout\tfile=toc.tmp
   \def\writetoca##1{{\edef\next{\write\tfile{\noindent ##1
   \string\leaderfill {\noexpand\number\pageno} \par}}\next}}}

%
\catcode`\@=12 
%
%
\font\abssl=cmsl10 scaled 833
\font\absrm=cmr10 scaled 833 \font\absrms=cmr7 scaled  833
\font\absrmss=cmr5 scaled  833 \font\absi=cmmi10 scaled  833
\font\absis=cmmi7 scaled  833 \font\absiss=cmmi5 scaled  833
\font\abssy=cmsy10 scaled  833 \font\abssys=cmsy7 scaled  833
\font\abssyss=cmsy5 scaled  833 \font\absbf=cmbx10 scaled 833
\skewchar\absi='177 \skewchar\absis='177 \skewchar\absiss='177
\skewchar\abssy='60 \skewchar\abssys='60 \skewchar\abssyss='60
\def\abstractfont{\def\rm{\fam0\absrm}
\textfont0=\absrm \scriptfont0=\absrms \scriptscriptfont0=\absrmss
\textfont1=\absi \scriptfont1=\absis \scriptscriptfont1=\absiss
\textfont2=\abssy \scriptfont2=\abssys \scriptscriptfont2=\abssyss
\textfont\itfam=\absi \def\it{\fam\itfam\absi}
\textfont\slfam=\abssl \def\sl{\fam\slfam\abssl}
\textfont\bffam=\absbf \def\bf{\fam\bffam\absbf}\rm}
\font\ftsl=cmsl10 scaled 833
\font\ftrm=cmr10 scaled 833 \font\ftrms=cmr7 scaled  833
\font\ftrmss=cmr5 scaled  833 \font\fti=cmmi10 scaled  833
\font\ftis=cmmi7 scaled  833 \font\ftiss=cmmi5 scaled  833
\font\ftsy=cmsy10 scaled  833 \font\ftsys=cmsy7 scaled  833
\font\ftsyss=cmsy5 scaled  833 \font\ftbf=cmbx10 scaled 833
\skewchar\fti='177 \skewchar\ftis='177 \skewchar\ftiss='177
\skewchar\ftsy='60 \skewchar\ftsys='60 \skewchar\ftsyss='60
\def\footnotefont{\def\rm{\fam0\ftrm}
\textfont0=\ftrm \scriptfont0=\ftrms \scriptscriptfont0=\ftrmss
\textfont1=\fti \scriptfont1=\ftis \scriptscriptfont1=\ftiss
\textfont2=\ftsy \scriptfont2=\ftsys \scriptscriptfont2=\ftsyss
\textfont\itfam=\fti \def\it{\fam\itfam\fti}%
\textfont\slfam=\ftsl \def\sl{\fam\slfam\ftsl}%
\textfont\bffam=\ftbf \def\bf{\fam\bffam\ftbf}\rm}
\font\ninerm=cmr9 \font\sixrm=cmr6 \font\ninei=cmmi9 \font\sixi=cmmi6
\font\ninesy=cmsy9 \font\sixsy=cmsy6 \font\ninebf=cmbx9
\font\nineit=cmti9 \font\ninesl=cmsl9 \skewchar\ninei='177
\skewchar\sixi='177 \skewchar\ninesy='60 \skewchar\sixsy='60
\def\ninepoint{\def\rm{\fam0\ninerm}
\textfont0=\ninerm \scriptfont0=\sixrm \scriptscriptfont0=\fiverm
\textfont1=\ninei \scriptfont1=\sixi \scriptscriptfont1=\fivei
\textfont2=\ninesy \scriptfont2=\sixsy \scriptscriptfont2=\fivesy
\textfont\itfam=\ninei \def\it{\fam\itfam\nineit}\def\sl{\fam\slfam\ninesl}%
\textfont\bffam=\ninebf \def\bf{\fam\bffam\ninebf}\rm}
%
%

\vsize=7.0truein
\hsize=4.7truein
\baselineskip 12truept plus 0.5truept minus 0.5truept
\hoffset=0.5truein
\voffset=0.5truein
\def\1{\;1\!\!\!\! 1\;}

\def\tr{\,{\hbox{tr}}\,}

\def\epm#1#2{\hbox{${+#1}\atop {-#2}$}}

\def\gsim{\mathrel{\rlap{\lower4pt\hbox{\hskip1pt$\sim$}}
    \raise1pt\hbox{$>$}}}         

\def\frac#1#2{{{#1}\over {#2}}}
\def\half{\hbox{${1\over 2}$}}

\def\tr{{\rm tr}}

\def\MS{\hbox{$\overline{\rm MS}$}}

\catcode`@=11 
\def\slash#1{\mathord{\mathpalette\c@ncel#1}}
 \def\c@ncel#1#2{\ooalign{$\hfil#1\mkern1mu/\hfil$\crcr$#1#2$}}
\def\lsim{\mathrel{\mathpalette\@versim<}}
\def\gsim{\mathrel{\mathpalette\@versim>}}
 \def\@versim#1#2{\lower0.2ex\vbox{\baselineskip\z@skip\lineskip\z@skip
       \lineskiplimit\z@\ialign{$\m@th#1\hfil##$\crcr#2\crcr\sim\crcr}}}
\catcode`@=12 

\def\PR{{\it Phys.~Rev.~}}
\def\PRL{{\it Phys.~Rev.~Lett.~}}
\def\NP{{\it Nucl.~Phys.~}}

\def\PL{{\it Phys.~Lett.~}}

\def\ZP{{\it Zeit.~Phys.~}}

\def\vol#1{{\bf #1}}\def\vyp#1#2#3{\vol{#1}, #3 (#2)}


\tolerance=10000
\hfuzz=5pt
\pageno=0\nopagenumbers\tolerance=10000\hfuzz=5pt
\line{\hfill {\tt hep-ph/9608399}}
\line{\hfill Edinburgh 96/23}
\line{\hfill DFTT 37/96}
\vskip 12pt
\centerline{\bf POLARIZED PARTONS}
\centerline{\bf AT NEXT-TO-LEADING ORDER}
\vskip 12pt
\centerline{Stefano Forte}
\vskip 6pt
\centerline{\it INFN, Sezione di Torino}
\centerline{\it via P. Giuria 1, I-10125 Torino, Italy}
\vskip 10pt
\centerline{Richard D. Ball\footnote*{\footnotefont
Royal Society University Research Fellow}}
\vskip 6pt
\centerline{\it Department of Physics and Astronomy}
\centerline{\it University of Edinburgh, EH9 3JZ, Scotland}
\vskip 10 pt
\centerline{Giovanni Ridolfi}
\vskip 6pt
\centerline{\it INFN, Sezione di Genova}
\centerline{\it via Dodecaneso 33, I-16146 Genova, Italy}
\vskip 24pt
{\narrower\baselineskip 10pt
\centerline{\bf Abstract}
\medskip\noindent
We discuss the determination of polarized parton distributions
from a NLO analysis of recent experimental data. We extract the first
moment of the polarized quark and gluon distribution and assess the
corresponding uncertainties, with special regard to those related
to scheme dependence.
\smallskip}
\vskip 48pt
\centerline{Talk given at {\it DIS96}, Rome, April 1996}
\medskip
\centerline{\it to be published in the proceedings}
\vfill
\line{August 1996\hfill}
\eject
\footline={\hss\tenrm\folio\hss}
\centerline{\bf POLARIZED PARTONS}
\centerline{\bf AT NEXT-TO-LEADING ORDER}
\bigskip\bigskip
{\ninepoint
\centerline{STEFANO~FORTE}
\smallskip
\centerline{\it INFN, Sezione di Torino}
\centerline{\it via P. Giuria 1, I-10125 Torino, Italy}
\medskip
\centerline{RICHARD D.~BALL}
\smallskip
\centerline{\it Department of Physics and Astronomy,}
\centerline{\it University of Edinburgh, EH9 3JZ, Scotland}
\medskip
\centerline{GIOVANNI RIDOLFI}
\smallskip
\centerline{\it INFN, Sezione di Genova}
\centerline{\it via Dodecaneso 33, I-16146 Genova, Italy}
\smallskip

}
\bigskip
{\abstractfont\baselineskip 9 pt
\advance\leftskip by 36truept\advance\rightskip by 36truept\noindent
We discuss the determination of polarized parton distributions
from a NLO analysis of recent experimental data. We extract the first
moment of the polarized quark and gluon distribution and assess the
corresponding uncertainties, with special regard to those
related to scheme dependence.
\smallskip}

\baselineskip 12pt plus 0.5pt minus 0.5pt
\bigskip\bigskip
\nref\smc{SMC Collaboration,
D.~Adams et al., \PL\vyp{B329}{1994}{399}; \PL\vyp{B357}{1995}{248}.}
\nref\slac{E143 Collaboration, K.~Abe et al., \PRL\vyp{74}{1995}{346};
\PRL\vyp{75}{1995}{25}.}
\nref\slacb{E143 Collaboration, K.~Abe et al., \PL\vyp{B364}{1995}{61}.}
\nref\neer{R.~Mertig and W.L.~van~Neerven, \ZP\vyp{C70}{1996}{637}.}
\nref\vogel{W.~Vogelsang, \PR\vyp{D54}{1996}{2023}.}

The theoretical and experimental information on inclusive
polarized deep-inelastic scattering has considerably
improved recently. On the one hand, the polarized structure function $g_1$
has now been measured over a reasonably wide range in $x$ at several
scales, both for proton and deuterium
targets.\refs{\smc-\slacb} On the other hand, a complete
determination of the two-loop polarized anomalous dimensions is now
available.\refs{\neer,\vogel} This makes a fully-fledged NLO
analysis of polarized parton distributions both feasible and justified.
Such an analysis is necessary
in order to extract the physically relevant matrix elements
of quark and gluon operators from the measured $g_1(x,Q^2)$.

The simplest leading-twist polarized observables are the nucleon
matrix elements of the quark axial currents
$j^\mu_{5,i}=\bar \psi_i \gamma_\mu\gamma_5 \psi_i$: for a
target of mass $M$ and spin $s^\mu$ the matrix elements
$\langle s |j^\mu_{5,i} | s \rangle = M a_i(Q^2) s^\mu$ are related
to the first moment of
$\Gamma_1(Q^2)\equiv\int_0^1\!dx\,g_1(x,Q^2)$ by
\eqn\fmom{
\Gamma_1(Q^2)=
\hbox{$\half\sum_{i=1}^{n_f}$} e^2_i C_i
a_i=\half\left[\langle e^2\rangle C_{\rm S}(Q^2)a_0(Q^2)+
C_{\rm NS}(Q^2)a_{\rm NS}(Q^2)
\right],}
where the average quark electric charge is
$\langle e^2\rangle={1\over n_f}\sum_{i=1}^{n_f} e^2_i$,
the singlet and nonsinglet axial charges are respectively
$a_0 =\sum_{i=1}^{n_f} a_i$ and
$a^{\rm NS}=\sum_{i=1}^{n_f}
\left(e_i^2-\langle e^2\rangle\right) a_i$, and $C(Q^2)$
are perturbatively computable coefficient
functions, which at NLO are given by $C_{\rm S}(Q^2)=C_{\rm NS}(Q^2)
=1-{\alpha_s\over\pi}+O(\alpha_s^2)$.\ref\koda{
J.~Kodaira et al., \PR\vyp{D20}{1979}{627}.}
While the nonsinglet axial charge $a_{\rm NS}(Q^2)$ only depends on scale
due to heavy quark thresholds and is otherwise determined by SU(3) symmetry
in terms of hyperon $\beta$ decay constants, the singlet charge measures a
nontrivial combination of the first moments of the polarized quark and gluon
distributions:
\eqn\pdfmom
{C_{\rm S}(Q^2)a_0(Q^2)=C_q(1,Q^2) \Delta \Sigma(1,Q^2)+
C_g(1,Q^2) \Delta g(1,Q^2)}
where $\Delta \Sigma(N,Q^2)=\sum_{i=1}^{n_f}
\int_0^1\!dx\, x^{N-1} [\Delta q_i+\Delta \bar q_i](x,Q^2)$ is the
moment-space polarized singlet quark distribution, other Mellin
moments are defined similarly, $C_q(1,Q^2)=1+ O(\alpha_s(Q^2)) $
and $C_g(1,Q^2)=O(\alpha_s(Q^2))$.

\nref\bfr{R.D.~Ball,
S.~Forte and G.~Ridolfi, \NP\vyp{B444}{1995} {287}.}\nref\bfra{R.D.~Ball,
S.~Forte and G.~Ridolfi, \PL\vyp{B378} {1996} {255}.}
A direct determination of $a_0(Q^2)$ would be possible only if $g_1$
were known for all $x$ at one scale; since this is not the case we
can only determine it indirectly through eq.~\pdfmom\
from quark and gluon distributions,
the primary quantities which enter the QCD evolution
equations. All the available experimental information on $g_1(x,Q^2)$ is then
summarized by constructing a set of polarized parton distributions (gluon and
quark singlet and nonsinglet)
at a reference scale $Q_0$, from which $g_1$ can be obtained for all $x$ and
$Q^2$ by solving the evolution equations. We parametrize these distributions
as $\Delta f(x, Q_0)=N_f\eta_f
x^{\alpha_f} (1-x)^{\beta_f}
(1+a_fx)$ (with $N_f$ chosen in such a way that the first moment of
$\Delta f$ equals $\eta_f$).  We evolve these distributions to all the
points in the $(x,Q^2)$ plane where $g_1(x,Q^2)$ has been measured,
and thereby determine the best fit values of the parameters
(further details of the fitting procedure are given in
refs.~\xref\bfr,~\xref\bfra). We use the NLO form of
anomalous dimensions and coefficient functions, determined in the \MS\
renormalization scheme.

The choice of factorization scheme is particularly
subtle in the polarized case because of the extra ambiguity related
to the prescription used to define the $\gamma_5$ matrix in
dimensional regularization; without such a prescription the \MS\
scheme is not fully defined.
We will use the so-called Adler-Bardeen scheme,\cite\bfra\ which
is defined in terms of a specific factorization scheme change
on the singlet anomalous dimensions and coefficient
functions of refs.~\xref\neer,\xref\vogel\ (which are in turn obtained
from \MS\ factorization and the 't~Hooft-Veltman prescription for
$\gamma_5$ by a scheme change which enforces the conservation of the
nonsinglet axial current).
The AB scheme is constructed by imposing conditions on first moments
of singlet anomalous dimensions and coefficient functions, whose
factorization scheme dependence is especially delicate.

To understand this, recall that $a_0$ depends on
$Q^2$ because the singlet axial current, unlike its nonsinglet counterpart,
is not conserved in the chiral limit in the quantized theory due to
the axial anomaly: $\partial_\mu j^\mu_5
=n_f{\alpha_s\over2\pi}\tr\epsilon^{\mu\nu\rho\sigma}
F_{\mu\nu}F_{\rho\sigma}$. As a consequence, the current is
multiplicatively renormalized:
${d\over d t}  j^\mu_5=\gamma^{(5)}(\alpha_s) j^\mu_5$, with an
anomalous dimension
$\gamma^{(5)} =\gamma^{(5)}_{(2)}\alpha_s^2+O(\alpha_s^3)$
that starts at two loops. The anomaly equation then implies that the
(gluonic) anomaly operator mixes with the axial current according to
${d\over dt}n_f{\alpha_s\over2\pi}\tr\epsilon^{\mu\nu\rho\sigma}
F_{\mu\nu}F_{\rho\sigma}=\gamma^{(5)}(\alpha_s)\partial_\mu j^\mu_5$.
At the level of operator matrix elements, the current is parametrized
by $a_0(Q^2)$, and
the  partonic quantity  which mixes with it can be identified with the
first moment of the polarized gluon distribution:
\eqn\evglu{
\hbox{${{d}\over{d t}}$}\left[-n_f\hbox{${\alpha_s\over2\pi}$}
\Delta g(1, Q^2)\right]
=\gamma^{(5)}(\alpha_s) a_0(Q^2).}
This is possible because at LO $\alpha_s\Delta g(1, Q^2)$ is scale
independent.

Eq.~\evglu\ implies that the scale dependence of $\Delta g(1, Q^2)$
causes a strong scheme dependence of $\Delta \Sigma(1,Q^2)$: upon a
general scheme change \hbox{
$\big({{\Delta \Sigma^\prime}\atop{\Delta g^\prime}}\big)=
\big[\1+\alpha_s
{z_{qq}\,z_{qg}\choose z_{gq}\,z_{gg}}+O(\alpha_s^2)\big]
\big({{\Delta \Sigma}\atop{\Delta g}}\big)$},
and since $\alpha_s \Delta g(1, Q^2)$
is scale independent at LO it follows that $\Delta \Sigma^\prime-
\Delta \Sigma$ does not vanish asymptotically if $z_{qg}\not=0$. Because
at LO $\Delta\Sigma$ is scale-independent this implies that even asymptotically
the ambiguity on $\Delta \Sigma(1,Q^2)$ is
of the same order as its size, i.e. the definition
of $\Delta\Sigma(1,Q^2)$ is entirely ambiguous. This is not the case
for the gluon: even if we  redefine it by setting
$z_{gg}\not=0$ then $\Delta g^\prime(1,Q^2)-\Delta g(1,Q^2)$ is
asymptotically constant, but since in the same limit
$\Delta g(1,Q^2)$ diverges (as $\alpha_s^{-1}$) the relative scheme
ambiguity in $\Delta g(1,Q^2)$ vanishes asymptotically.

The definition of $\Delta \Sigma(1,Q^2)$ is thus intrinsically arbitrary
and can only be fixed on the basis of physical requirements.
A simple choice is based on the observation that since the quantities
on either side of eq.~\evglu\ evolve in the same way, their difference
is scale independent: we can then identify this scale-independent
combination with the singlet quark first moment
\eqn\singq{\Delta \Sigma(1)=a_0(Q^2)+n_f\hbox{${\alpha_s\over2\pi}$}
\Delta g(1, Q^2)}
 to all orders. This choice has
several advantages: a) only when $\Delta \Sigma(1)$ is scale
independent does the Ellis-Jaffe ansatz,
which would equate it to the scale-independent octet axial charge $a_8$,
aquire an unambiguous meaning beyond LO;\ref\alros{G.~Altarelli and G.G.~Ross,
\PL\vyp{B212}{1988}{391}.}
b) the polarized quark distribution can then be defined directly
in terms of a hard observable (such as a jet cross section),
by properly factorizing all soft contributions to the cross section
into the parton distributions;\ref\fact{R.D.~Carlitz,
J.C.~Collins and A.H.~Mueller, \PL \vyp{B214}{1988}{229} and in Proc. of
24th Rencontre de Moriond, T.~T.~Vanh, ed, (Fronti\`eres, Paris, 1989);
G.~Altarelli and B.~Lampe, \ZP \vyp{C47}{1990}{315};
W.~Vogelsang, \ZP \vyp{C50}{1991}{275}.}  c)
$\Delta \Sigma (1)$ eq.~\singq\ coincides with the matrix element
of the canonically defined quark helicity operator.\ref\sfanom{S.~Forte,
\NP\vyp{B331}{1990}{1}.}

The AB scheme choice exploits the fact that eq.~\evglu\ is
automatically satisfied with the factorization choice of
refs.~\xref\neer,\xref\vogel, and
enforces the definition eq.~\singq\ of the polarized
quark distribution. This still leaves the freedom to further redefine
the normalization of $\Delta g(1,Q^2)$ (i.e. to vary $z_{gg}$ in the scheme
change), which can be fixed entirely at NLO by requiring that eq.~\evglu\
be also true at NNLO: in fact, the Adler-Bardeen theorem\ref\adbar{S.~Adler
and W.~Bardeen, \PR\vyp{182}{1969}{1517};
M.~Bos, \NP \vyp{B404}{1993}{215}.}
implies
that  there exists a scheme where eq.~\evglu\ is valid to all orders.
Knowledge\ref\lar{S.~A.~Larin, {\it Phys}. {\it Lett}.
{\bf B334} (1994) 192.} of the $O(\alpha_s^3)$ contribution to
$\gamma^{(5)}$ then fixes entirely the scheme choice for the first
moments of parton distributions. As a consequence, in the
AB scheme the evolution of $a_0$, as  given by eq.~\evglu, is automatically
correct through NNLO when the parton distributions
are evolved at NLO. The factorization scheme for all other moments
is finally fixed by insisting that the full scheme change
from the scheme of refs.~\xref\neer,\xref\vogel\ to the AB scheme
is independent of $x$.

The growth of the first moment of the polarized gluon distribution
with scale implies sizable scaling violations for moments close to
the first, which evolve at LO by mixing with the gluon distribution.
Measuring these scaling violations then determines
the polarized gluon distribution. This
determination is only asymptotically scheme independent,
while at finite $Q^2$ it is  subject to a factorization scheme
ambiguity.\footnote*{\footnotefont\baselineskip 10 truept Note that
the Altarelli-Parisi equations allow a determination of all
moments of $\Delta g$ from scaling violations, including the first,
provided only that the input distributions are sufficiently smooth.
This is despite the fact
that the singlet first moment evolves multiplicatively, and thus
measurements at different scales determine only one linear combination
of $\Delta \Sigma(1)$ and $\Delta g(1)$. However, two functions
of $x$ whose moments are all identical, but whose first moments are
different, can only differ by a term proportional to $\delta(x)$.
Such terms are not experimentally measurable, even in principle, and
are thus by construction not included in the definition of parton
distributions.
The first moment $\Delta g(1)$ is thus uniquely determined by
analytic continuation from $\Delta g(N)$, $N\neq 1$. In practice
this analytic continuation is performed implicitly through the choice
of the form of the parametrization of the initial parton
distributions.
\vskip1pt}
On the other hand, large scaling violations driven by a large gluon
distribution also imply large scheme ambiguities, in particular in the
determination of $a_0$. For instance, it is clear from eq.~\pdfmom\ that
at NLO $a_0(Q^2)$ will be affected by a scheme ambiguity of order $\alpha_s^2
\Delta g$, which only decreases with scale as $\alpha_s$.

\nref\psx{R.D.~Ball, {\tt hep-ph/9511330}, in the proceedings of the
``International School on Nucleon Spin Structure'', Erice 1995.}
Fitting to the data of refs.\xref\smc,\xref\slac\ we find\cite\bfra
\eqn\salame{\eqalign{&a_0(10~{\rm GeV}^2)=0.14\pm 0.10\>\hbox{(exp.)}
\epm{0.12}{0.05}\>\hbox{(th.)}\cr
&\Delta \Sigma(1)
=0.5\pm0.1,\qquad\Delta g(1,10\> \hbox{GeV}^2 )
=3.0\pm 1.6.\cr}}
The largest source of experimental error comes from the uncertainty in
the behaviour of the partons in the unmeasured small $x$
region,\refs{\bfr,\bfra,\psx} which could
of course in principle lead to an infinite (either positive or
negative) contribution to first moments.
In ref.\xref\bfra\ it is constrained by our specific choice
of functional form for the initial parton distributions: assuming that
the small $x$ exponent $\alpha_f$ can be fitted from the data amounts to
assuming that the asymptotic small $x$ behaviour has already started
setting in in the measured region ($x\ge 0.003$). The data turn out to
determine reasonably well the small $x$ behaviour of the nonsinglet
($\alpha_{\rm NS}=-0.68\pm0.15$), but only very poorly that of the
singlet quark ($\alpha_{\rm S}=0.41\pm0.38$) and gluon
($\alpha_{\rm g}=-0.47\pm0.30$); this is then reflected in a sizable
statistical uncertainty in the first moments.
The bulk of the theoretical error is due to lack of knowledge of higher order
corrections, as manifested by scheme dependence, which
we estimate by varying the renormalization
and factorization scales by a factor of two around $Q^2$. Further
sources of error, which include the variation of the position of heavy
quark thresholds, the value of $\alpha_s$ and SU(2) and SU(3) breaking
(as reflected in the values of the nosinglet axial charges)
are negligible in comparison.

Since the error on the determination of the polarized gluon distribution
is dominated by the statistics, it can be improved when
more precise or abundant data in the presently measured region become
available: for instance, including
the data of ref.~\xref\slacb\ in the analysis leads to
$\Delta g(1,10\> \hbox{GeV}^2 )
=2.6\pm 1.3$, even though the statistical weight of these data is
relatively modest. The determination of the axial charge $a_0$, on
the other hand, is dominated by the systematics related to higher order
corrections, and the uncertainty in the small $x$ extrapolation.
The former could be improved by higher statistics data which provide
a more complete coverage of the $x-Q^2$ plane (allowing evaluations of
the first moment at essentially a single value of $Q^2$). However, to
obtain a significant improvement in the latter, data in the unexplored
small $x$ region will  be required.
A substantial improvement in the determination of the singlet axial charge
will thus only be accomplished by the next generation of experiments, such
as would be possible at HERA with a polarized proton
beam.\ref\jech{J.~Lichtenstadt, these proceedings; R.~D.~Ball et al.,
in the proceedings of the workshop on ``Future physics at HERA''.}
\smallskip
{\bf Acknowledgments:}
We thank A.~Deshpande and J.~Lichtenstadt for interesting conversations
on their data, and G.~Altarelli for discussions.

\noindent

\immediate\closeout\rfile\writestoppt
\bigskip
\noindent{{\bf References}}\smallskip{\frenchspacing%
\parindent=20pt
\ninepoint\baselineskip=11pt
\escapechar=` \input refs.tmp\vfill\eject}\nonfrenchspacing

\end